# Using the Fermilab Proton Source for a Muon to Electron Conversion Experiment

C. Ankenbrandt, D. Bogert, F. DeJongh, S.Geer, D. McGinnis, D. Neuffer,
M. Popovic, E. Prebys

*Fermi National Accelerator Laboratory, PO Box 500, Batavia, IL 60510*

Abstract

The Fermilab proton source is capable of providing 8 GeV protons for both the future long-baseline neutrino program (NuMI), and for a new program of low energy muon experiments. In particular, if the 8 GeV protons are rebunched and then slowly extracted into an external beamline, the resulting proton beam would be suitable for a muon-to-electron conversion experiment designed to improve on the existing sensitivity by three orders of magnitude. We describe a scheme for the required beam manipulations. The scheme uses the Accumulator for momentum stacking, and the Debuncher for bunching and slow extraction. This would permit simultaneous operation of the muon program with the future NuMI program, delivering $10^{20}$ protons per year at 8 GeV for the muon program at the cost of a modest (~10%) reduction in the protons available to the neutrino program.

## 1. INTRODUCTION

Searches for lepton flavor violation in rare muon processes can probe new physics at mass scales comparable to, and beyond, the scales probed at high energy colliders [1]. A new generation of these experiments would complement the future LHC and ILC programs, and help elucidate the discoveries made at the energy frontier. Rare muon experiments require a high intensity muon source, which in turn requires a high intensity proton source to produce the pions that then decay into muons. We propose a scheme to develop the existing Fermilab proton source so that it can produce a primary proton beam suitable for a new generation of rare muon experiments. Specifically, the scheme would produce a proton beam of sufficient intensity and appropriate bunch structure for a next-generation experiment searching for muon-to-electron conversion in the field of a nucleus ($\mu^-N \rightarrow e^-N$). This beam could be run in parallel with the future Fermilab neutrino program (SNuMI [2]), and would provide $10^{20}$ protons per year at 8 GeV for the muon program at the cost of a modest 10% reduction in the protons available for SNuMI. The resulting proton source would enable, with 4 years of running, a MECO-like experiment [3] to improve the present sensitivity to $\mu^-N \rightarrow e^-N$ by three orders of magnitude.

## 2.    BEAM REQUIREMENTS

In a $\mu^-N \rightarrow e^-N$ experiment, the negative muons come to rest in a target, and are captured into an orbit around a target nucleus. They can then either decay, be radiatively captured by the nucleus, or do something more exciting (e.g. convert to an electron). The radiative capture lifetime depends upon the target nucleus. For aluminum the capture time is $\tau_{Al}$ = 1.1 $\mu$s [3]. If the muon converts into an electron before it decays or is radiatively captured, it will produce a mono-energetic electron recoiling against the nucleus. The experimental signature for $\mu^-N \rightarrow e^-N$ is, therefore, a single unaccompanied electron with a characteristic energy which, for an Al target, is 105.1 MeV.

In a MECO-like $\mu^-N \rightarrow e^-N$ experiment, after a muon bunch arrives the experiment waits a few hundred nanoseconds before data taking. This allows all the beam particles and backgrounds not stopped in the target to exit the detector region. Data is then taken until the next beam bunch arrives. The experiment therefore requires a bunched beam with the following characteristics:

i) The <u>bunch length</u> must be short compared to the capture lifetime. Data can only be taken between bunches since the presence of incident energetic particles can cause backgrounds in the detector. If the bunch is long compared to the muon radiative capture lifetime, most of the stopped muons will have disappeared before the experiment can start looking for $\mu^-N \rightarrow e^-N$.

ii) The <u>bunch separation</u> must be comparable to, or longer than, the radiative capture lifetime. We do not want the next bunch to arrive (which will stop data taking) while there are still a useful number of stopped muons surviving within the target.

iii)   The <u>bunch separation</u> must not be very long compared to the capture

lifetime. For a given number of incident muons, we would like to minimize the instantaneous intensity, and hence maximize the number of muon bunches. Therefore, we would like the minimum bunch separation consistent with criteria (ii).

iv) <u>Between muon bunches</u> there must be no incident beam particles. The fraction of out-of-time beam particles (extinction factor) must be $10^{-9}$ or less for a MECO-like experiment if backgrounds are not to reduce its advertised sensitivity.

## 3. SNuMI PHASE 2 REPRISE

Once the Tevatron Collider program is completed there are three storage rings that become available for other uses: (i) The Recycler, a 3319m circumference ring in the MI enclosure; (ii) The Accumulator, a 474m circumference ring currently used to accumulate antiprotons; (iii) The Debuncher, a 504m circumference ring that shares the enclosure with the Accumulator, and is currently used to collect and debunch antiprotons. In the future it is proposed to use the Recycler and the Accumulator to increase the number of 8 GeV protons that can be injected into the MI per MI cycle, and hence the number of protons available for the MI neutrino program NuMI. This upgrade to the NuMI facility (SNuMI [4]) is expected to occur in two phases. We propose extending the second phase (SNuMI Phase 2) to include additional modifications that would support a muon program that could run in parallel with the SNuMI program.

**3.1 SNuMI Phase 1**: The MI injection time is short compared to the time to ramp up and ramp down. To increase the number of protons per unit time available for the NuMI program requires decreasing the ramping time (and hence the MI cycle time) and/or increasing the number of protons that can be injected during the short injection time. At present the MI cycle is 22 Booster cycles long. Hence, only one out of 22 Booster batches is injected into the MI. In SNuMI phase 1, the Recycler is used to accumulate the Booster protons while the MI is ramping; 12 Booster batches are "boxcar" stacked in the Recycler during the MI cycle, and then injected into the MI. Hence 12 out of 22 Booster batches will be injected into the MI.

**3.2 SNuMI Phase 2**: In the second phase of the SNuMI upgrade, the number of Booster protons injected into the MI per cycle is further increased by using the Accumulator to "momentum stack" three Booster batches before they are transferred into the Recycler. This is repeated 6 times per MI cycle, the 6 super-batches being boxcar stacked in the Recycler. Thus, 18 out of 22 Booster batches will be injected into the MI. SNuMI phase 2 will require two new beamlines to be built (Fig. 3.1); AP-4 which will take the Booster protons to the Accumulator, and AP-5 which will take the protons in the Accumulator to the Recycler. AP-4 parallels the historic route of the original AP-4 abandoned and destroyed when the Booster to MI 8 GeV line was constructed in 1996. AP-5 is a new line that takes extracted protons from under the AP-10 Building, around a 100 degree arc, back into the 8 GeV line just upstream of the access hatch near 817.

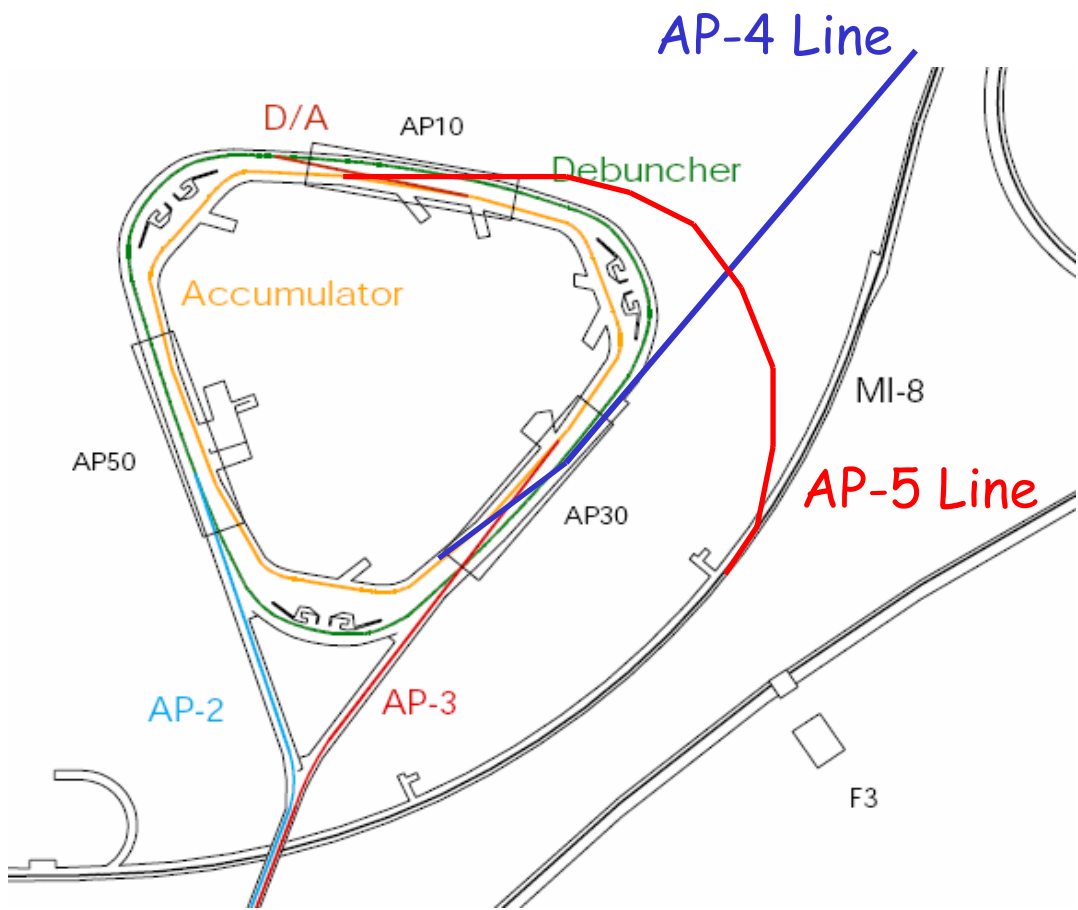

Figure 3.1: Antiproton source (Debuncher and Accumulator) schematic showing the AP4 (Booster to Accumulator) and AP5 (Accumulator to Recycler) lines to be built for SNuMI phase 2.

## 4.    SCHEME FOR A MUON PROGRAM

A $\mu^-N \rightarrow e^-N$ experiment requires a bunch structure which is very different from the bunch structure produced by the Fermilab Booster.  We propose to modify the SNuMI Phase 2 scheme to permit the Fermilab proton source to provide beam with the appropriate bunch structures for both the neutrino program and for a muon program. The modified scheme would use, within a single MI cycle, the Accumulator to momentum stack protons for the neutrino program (18 Booster batches) and then for the muon program (4 Booster batches). The Debuncher would then be used, for the muon program protons, to form a single bunch within the ring with a bunch length much shorter than the ring circumference (which is 1.69 μs). The Accumulator would be connected to the Debuncher using an A-D line which would be the present D-A line, relocated and reversed. The injection septum would be located at downstream D50, and would use, relocated, the present extraction septum at upstream D10. The bunched beam would be slowly extracted to an external beamline, producing a train of bunches with bunch lengths short (compared to 1.69 μs) and bunch separations of 1.69 μs.  The estimated time required to rebunch the beam in the Debuncher is much shorter than the MI cycle time, and hence slow extraction can take place over almost the entire MI cycle. This will produce an ideal bunch structure for a MECO-like experiment.

Table 4.1:  Debuncher Ring Parameters

| Parameter | Symbol | Value |
|---|---|---|
| Circumference | $C=2\pi R$ | 504m |
| Beam Momentum | P | 8.89 GeV/c |
| Transition | $\gamma_t$ | 7.52 |
| Betatron functions (max) | $\beta_x, \beta_y, \eta$ | 19.8, 17, 2.2m |
| Tunes | $\nu_x, \nu_y$ | 9.66, 9.76 |
| Period | $C/\beta c$ | 1690ns |

To illustrate how the modified scheme works to provide beam for both neutrino and muon programs, a timing diagram is shown in Fig. 4.1. For the representative example shown, the MI cycle is 22 Booster cycles (1.467s) long. Within each MI cycle, 18 batches are used for the neutrino program, and 4 batches are used for the muon program. The first 3 batches for the neutrino program are momentum stacked in the Accumulator, and then transferred to the Recycler. This is repeated 6 times, box-car stacking each triple-batch in the recycler. The next 4 Booster batches are then momentum stacked in the Accumulator, and transferred to the Debuncher for the muon program. Allowing 0.1s for rebunching and beam cleanup in the Debuncher, the protons are then slowly extracted into an external beamline for 1.367s, providing a bunched beam for the muon program. Assuming $4.6 \times 10^{12}$ protons per Booster batch, this scheme provides $56 \times 10^{12}$ protons/sec for the NuMI program, and $12.5 \times 10^{12}$ protons/sec for the muon program. Note that, in the absence of a muon program, the MI cycle time could be reduced to 1.33s, yielding 18 Booster batches per 1.33s for the neutrino program = $62 \times 10^{12}$ protons/sec. Hence, the additional muon program would reduce the protons available for the neutrino program by only ~10%.

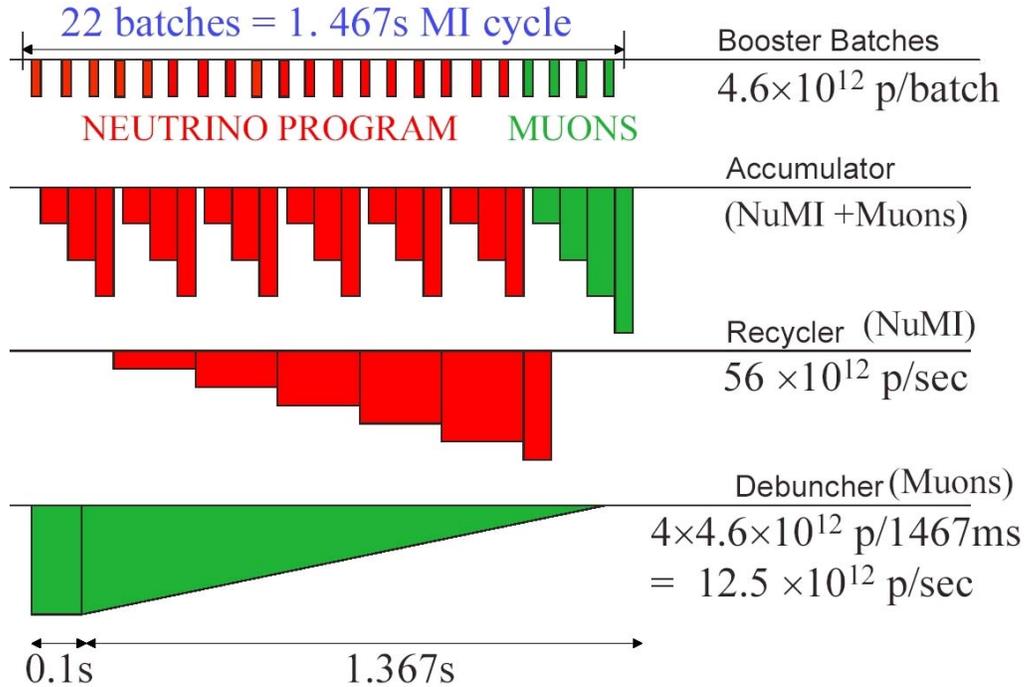

**Figure. 4.1:** Timing diagram showing machine usage during one MI cycle.

## 5. REBUNCHING IN THE DEBUNCHER

The proton beam is debunched in the Accumulator so that it fills the circumference (474m or 1600ns). The rf manipulation during this process also decreases the energy spread. After accumulation of 4 Booster batches the full emittance is estimated to be 84×0.38 = 32 eV-s, which implies an energy spread of 20 MeV (full width) in the fully debunched beam. An extraction gap of >45ns is introduced into the circumference-filling bunch (by a barrier bucket or harmonic rf) so that a fast kicker can be switched on, and the long bunch extracted into the Debuncher. The Debuncher circumference is ~30m greater than the Accumulator circumference, which increases the initial gap in the circulating beam to >145ns.

The $\mu^-N \rightarrow e^-N$ experiment requires that the beam be bunched to a small fraction of the ring circumference. This means the rf must increase the initial gap and hence compress the bunch into a short length. We have chosen a full-length of $\Delta\tau_B$<200ns as a goal in the bunching, which should meet the $\mu^-N \rightarrow e^-N$ experiment requirements. The bunch must be held to that length throughout the extraction cycle, with minimal leakage into the inter-bunch gap that could lead to mistimed extracted beam.

## 5.1 Rf Bunching

Scenarios for rf bunching in the Debuncher have been explored, initially by considering only the longitudinal motion of the beam. The equations of motion are:

$$\frac{d\phi}{dn} \cong \frac{2\pi}{\beta^2\gamma}\left(\frac{1}{\gamma^2} - \frac{1}{\gamma_t^2}\right)\frac{\Delta E}{mc^2} = \frac{2\pi}{\beta^2\gamma}\alpha_p\,\frac{\Delta E}{mc^2}$$

$$\frac{d\Delta E}{dn} = eV_{RF}(\phi)$$

where n is the number of turns, $\phi$ is the particle phase (longitudinal position) and $\Delta E$ is the particle energy offset from the reference energy. $V_{rf}(\phi)$ is the rf voltage, $m$ is the proton mass, $\beta$=v/c and $\gamma = (1-\beta^2)^{-1/2}$ are the usual kinematic factors, and $\alpha_p = 1/\gamma^2 - 1/\gamma_t^2$ is the momentum compaction factor. In the Debuncher, $\gamma$=9.52 and $\gamma_t$=7.6, so $\alpha_p$ =-0.006. The small value of $\alpha_p$ means the motion is nearly isochronous, which implies that the longitudinal motion is relatively slow. That means that only a small amount of $V_{rf}$ will be needed in the rebunching, but that bunching may be relatively slow.

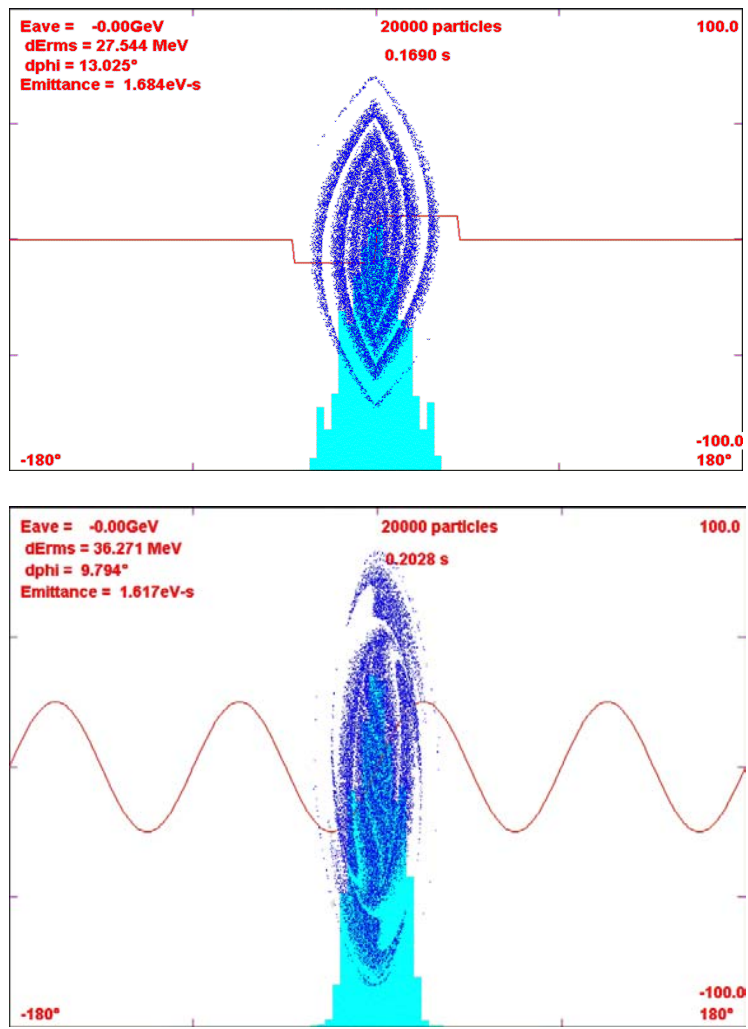

Figure 5.1: 1-D simulation of bunch compression. A: Longitudinal distribution after 0.169s of barrier bucket compression. B: Longitudinal distribution after further h=4 bunching (Emittance numbers on plots are rms emittances; full emittances are a factor of $6\pi$ larger).

## 5.2 Barrier Bucket rf.

A simple model for bunching is based on the procedure used in the Recycler, which uses barrier bucket rf waveforms to bunch and debunch antiproton beams. In an initial example we can grow a square wave potential at two ends of the bunch and then move the phases of that square wave together until the two ends of the potential meet. In the present example we considered square waves of + and − 20 kV amplitude, with 40° phase widths (~180ns). The bunching must be slow enough to avoid phase space dilution; in the initial example the compression occurs over 0.17s (~10000 turns). At that point a single harmonic h=4 (2.38MHz) rf system is imposed at 25kV, with amplitude increased to 75kV over 0.03s. This rf is used to maintain the beam in a short bunch structure over the remainder of the 1.5s cycle (~1.3s).

The h=4 rf system was used since it is similar in voltage and frequency to an rf system that is currently used in the MI for Tevatron bunch coalescing (that rf will not be needed after 2010). This system bunches the beam to an rms bunch length of ~40ns, with the entire beam held within a ~200ns full width. The energy full width of the beam is increased to ~±100MeV. Our initial simulation indicates that phase-space dilution is limited to < ~20%. The 1-D simulation results of the model system are presented in Fig. 5.1.

### 5.3 Multi-harmonic rf buncher

The barrier bucket approach requires the use of a low-Q rf system, which is inefficient, may be somewhat expensive, and in practice may produce rf waveforms that are less ideal than assumed in our initial model. We therefore also consider an alternative bunching system consisting of a combination of low-harmonic rf sub-systems, which could be obtained using high-Q fixed-frequency cavities. In an initial example, we consider a combination of h=1, 2, 3, 4 rf cavities with maximum voltages of 30, 15, 10 and 7.5kV respectively, and combined in order to obtain an approximately linear rf wave form. In our simulation, these cavities are ramped from 0 to full strength over ~0.055s. As in the previous example, the 4th-harmonic rf is ramped up to 75kV to hold the beam for the remainder of the 1.5s extraction cycle. Simulation results are shown in Fig. 5.2. The final beam state is similar to the barrier bucket example, with an rms bunch length of ~40 ns and an energy full width < ~±100MeV. The finite number of harmonics resulted in a slightly more dilute phase space than a more idealized linear or barrier bucket rf would obtain; however, a more complete optimization and a more accurate evaluation of barrier bucket nonlinearity may remove the small difference. The bunching did occur in a significantly shorter time than the above barrier bucket example, however.

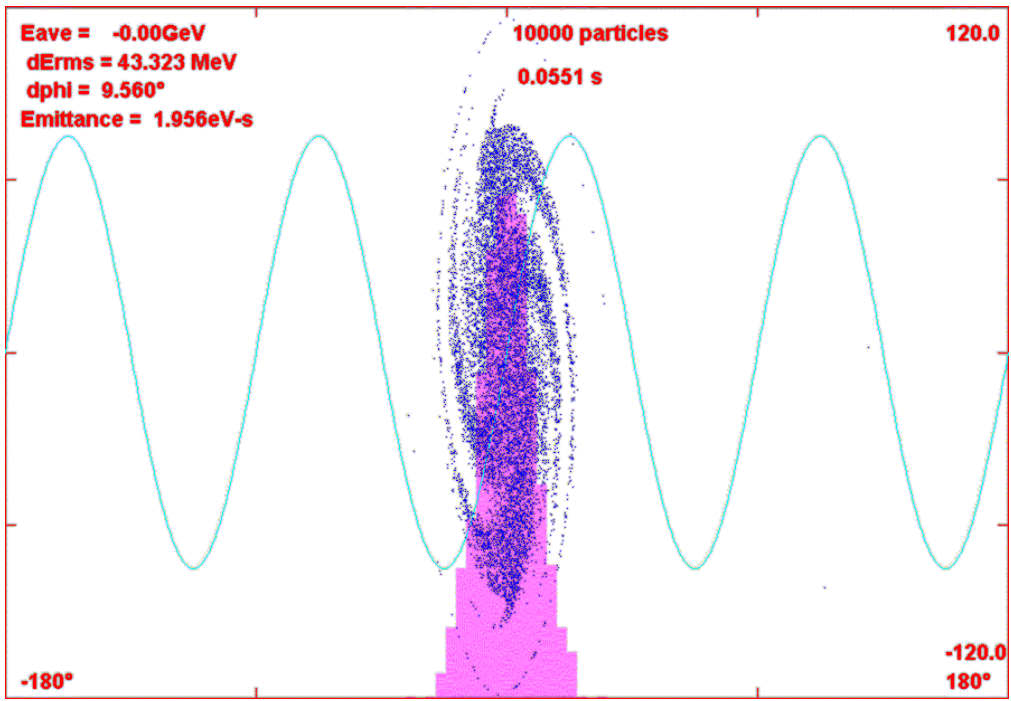

Figure 5.2. Beam after h=1, 2, 3, 4 adiabatic bunching with h=4 bunch compression; beam properties are similar to the previous barrier-bucket example but with somewhat more phase-space dilution.

## 5.4 Rebunching Discussion

Further study of instability intensity limitations is needed. The compressed beam has an enhanced space charge tune shift. A formula for that tune shift is:

$$\delta\nu \cong \frac{r_p N}{4\pi\beta\gamma^2 B_F \varepsilon_{N,rms}},$$

where $B_F$ is the bunching factor. With $B_F = 0.06$, N= $1.5\times10^{13}$, $\varepsilon_{Fermi} = 6\pi\varepsilon_{N,rms} = 20\pi$mm-mr, we obtain $\delta\nu \cong 0.1$, which should be acceptable.

The energy spread of $\pm100$MeV corresponds to $\delta p/p = \pm1\%$, which is less than the full acceptance of the Debuncher, which was designed to be $\pm2\%$. It is therefore possible to compress the beam to shorter lengths, with more rf and bunching time, but the additional reduction would be less than a factor of 2. The minimum bunch length is thus > 100ns. Some further study will be needed to determine whether this is adequate.

## 5.5 PRISM/PRIME compatible bunching mode

While this initial study is for a MECO-like experiment, a different approach is being developed by the PRISM/PRIME collaboration. In that approach, the proton beam is formed into short bunches, and single-bunch extracted onto a target to produce short $\pi\to\mu$ bunches. The short $\mu$ bunches are phase-energy rotated in the PRISM ring to small energy-spread bunches which are then single bunch extracted onto the PRIME target/detector. The phase-energy rotation requires bunches with $\sigma_\tau < 5$ns, roughly an order of magnitude shorter than the single intense bunch discussed above. However, the Debuncher proton beam could be bunched into a number of short bunches, and then be single turn extracted for a PRISM/PRIME-like experiment. As an example we bunch on harmonic h=12 (~7.14 MHz). The beam is initially adiabatically bunched, with an rf that increases to 10 kV over ~0.04s. The rf voltage is then increased to 200kV, and the bunches are phase-energy rotated in 0.001s to short $\delta\tau$ ($\sigma_\tau = 4$ns). Note that this is, qualitatively, the inverse of the present Debuncher cycle which takes a string of short antiproton bunches (small $\delta\tau$, large $\delta E$), fast-rotates them to small $\delta E$, and adiabatically debunches to small $\delta E$ contunous $\delta\tau$ for transfer to the accumulator. To maintain the short bunch length for later bunch extractions a higher- harmonic, high voltage rf system would be used. An h=36 or 48 system with $V_{rf} \cong 1$MV would be needed. Alternatively, the beam could be rotated to long bunches (~0.001s), held at 10kV, and rerotated at 200kV for short bunch extractions. In the baseline SNUMI-compatible mode, this produces ~12 bunches of protons, with ~$1.25\times10^{12}$ p/bunch every 1.5s, or ~$10^{20}$ protons/Snowmass-year. This is an order of magnitude less than desired by PRISM/PRIME. If not shared with SNUMI, this could be increased by ~4×, which may still not be enough. An eventual "proton driver" booster upgrade would reach that level, however.

## 6.    SLOW EXTRACTION

Th $\mu^-N \to e^-N$ experiment requires slow extraction to generate the required beam structure. Beam is rebunched into a single short bunch, and then slowly extracted over the remainder of the MI cycle. This results in a train of short, relatively low intensity pulses separated by the 1.69 µsec revolution time of the Debuncher.

The techniques of slow extraction are well established. The tune of the beam is moved near a harmonic resonance. That resonance is excited and a separatrix created in phase space. The tune is then swept toward that resonance and beam begins to "fall out" to higher amplitudes as it crosses the separatrix. The high amplitude beam is given an angular kick with a septum, followed by an extraction Lambertson roughly 90° later in phase. The beam fallout rate is tuned so that the amplitude increases to fill the aperture of the extraction septum over $N$ turns, where $N$ is the order of the resonance. In the ideal case, the inefficiency is entirely due to the beam striking the septum plane. For this reason, electrostatic septa are typically employed, which use a plane of small diameter wires to create the extraction field.

In practice, most slow extraction schemes have involved a third order resonance. This has the advantage of being analytically straightforward, and because it is inherently nonlinear, a separatrix is naturally produced. On the other hand, it is difficult to cleanly extract the last bit of beam. Fermilab has historically chosen to use half integer resonances. These are analytically more complex than third integer resonances, but have been thoroughly treated. They are complicated by the fact that the driving quadrupoles are inherently linear, so the beam would tend to go unstable simultaneously over the entire phase space area. A series of $0^{th}$ harmonic octupoles is used to introduce an amplitude dependent tune shift and thereby create a separatrix.

At this point, the details of the resonance are a secondary concern because the mechanics of the extraction are the same, namely an electrostatic septum followed by a Lambertson. The issues are that the lengths of the straight sections in the low dispersion regions are limited and that the beam will have to be bent rather sharply to clear the downstream quadrupole.

The initial proposal is to essentially mirror the existing injection scheme, with the extraction located at AP30 or AP50, depending on which experimental area is chosen. Assuming AP50 is chosen, the lattice and approximate layout is shown in Fig. 6.1. We first consider an extraction septum with approximately the specifications of the MI extraction septa: 80 kV over 1 cm, with a length of approximately 3 m. This would fit between Q403 and Q402 and would provide about 2.2 cm of horizontal displacement between Q501 and Q502. The extraction field over that straight region would need to be about 0.8 T to clear the Q502 quad. This field is well within the capabilities of the MI Lambertsons, but these only have an extraction channel of +- 5". A reasonable solution would be a short (~1 m) version of one of these, followed by a longer C magnet.

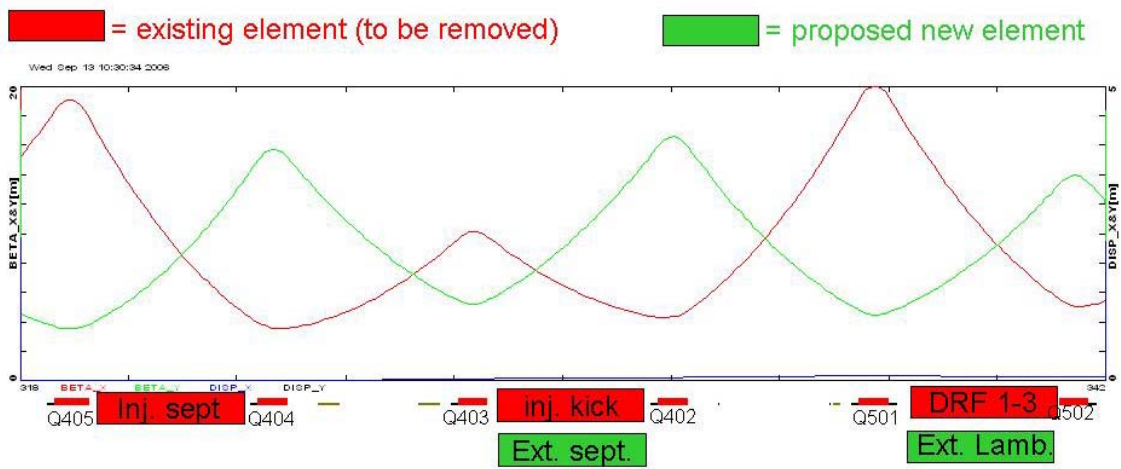

Figure 6.1 Lattice functions and approximate locations of extraction hardware. The labels in the top boxes indicate the existing hardware: injection septum, injection kicker and the DRF 1-3 cavity. These would all be removed and replaced with the elements shown in the lower boxes.

Beam loss is a significant worry. The best slow extraction schemes have an inefficiency of roughly 2-3%. In our proposed scheme, the beam power is roughly 20 kW, so about 500 W of beam will be lost. To set the scale, this is approximately equivalent to the entire uncontrolled beam loss in the Booster and it will be happening in an area with only a few feet of earth shielding. Shielding the loss will be complicated by the fact that much of it will scatter back into the Debuncher or down the extraction line. It's clear we must address this issue early in the design.

## 7. BEAM CLEANING

The extinction factor for this experiment is defined as the ratio of undesired, out-of-time protons to desired, in-time protons at the production target. The experiment calls for an extinction factor of one per billion or smaller. No single system is likely to achieve that factor, so it will be necessary to implement several measures. Protons that circumvent all the countermeasures will likely be the result of high-order processes that populate the tails of various distributions. It is difficult to anticipate of all the relevant high-order processes, let alone simulate them. Therefore, an experimental trial-and-error approach to achieve the required rejection is advisable. This will require instrumentation to monitor the rejection factor that is achieved at various stages of the beam delivery process. A real-time monitor of the final extinction factor would be ideal because it would provide continuous assurance that nothing unusual happened to allow an undesired burst of out-of-time protons through the system while the experiment is taking data.

The plan for this section is to go through the various beam-handling steps, noting opportunities along the way to suppress the delivery of undesired protons. It will turn out that many countermeasures can be conceived, and it seems likely that implementing all of them would be overkill. Some judgment will then be necessary to estimate the cost-effectiveness of various measures so that the most promising techniques can be implemented first. The trial-and-error process of implementing and testing systems sequentially will undoubtedly need substantial amounts of beam studies.

After the 8-GeV beam arrives in the Debuncher ring, the first beam-handling step is to form a single bunch that occupies a small fraction of the circumference. In that process it is likely that some beam, perhaps one to ten percent, will remain unbunched. It would be straightforward, albeit relatively expensive, to clear the gap by extracting the protons in the gap, via single-turn kickers through a septum, to a dump. This could be done periodically at a rate limited mainly by the achievable repetition rates of the kickers and any other pulsed extraction devices that may be used (e.g. the Booster 8-GeV extraction devices run at 15 Hz). These measures, taken together, should provide a rejection factor of several orders of magnitude. The unbunched particles that survive will mainly be those that coincide in time with the bunched beam when the kickers fire. That suggests that each abort cycle would provide about an order of magnitude additional rejection.

In addition to or instead of a single-turn abort system, particles in the gap could be removed by relatively modest time-varying fields in the ring. For example, a pulsed vertical dipole field or fields varying at the vertical fractional betatron frequency could drive the undesired particles in the gap toward the jaws of the collimation system that will probably be needed anyway to control losses in the ring. A similar system was proposed for the Brookhaven AGS for the MECO experiment.

The second beam-handling process is slow extraction. Since a detailed slow extraction design has not yet been worked out, it is important to realize that slow extraction can be implemented in ways that provide major rejection factors against unbunched beam. For example, the bunch could be slowly displaced in momentum, and the momentum offset together with nonzero chromaticity could move part of the beam toward the betatron frequency of the resonance that is used for extraction. An extraction system that uses time-varying fields seen only by the particles in the bunch is another possible strategy. One interesting example of that approach is to use one or more Tevatron electron lenses, which by will become surplus, to provide part of the tune shift that causes protons in the bunch to be extracted. It is worth noting that both of these methods might provide extracted bunches that are shorter in time than the bunch circulating in the ring.

The beam transport to the experiment provides opportunities for further enhancements of the extinction factor. One such possibility is a system of two dipoles separated by 180 degrees of betatron phase advance, each oscillating at the ring revolution frequency $\omega$ or half of that frequency. Such a two-bump would provide a displacement between the dipoles proportional to sin ($\omega$t), where t is the time measured from the arrival time of the center of the desired bunches. A collimation system between the two dipoles would then remove protons that are significantly out-of-time. Protons downstream of the second dipole would be undeflected.

Various opportunities exist to monitor the efficacy of the measures that are implemented to discriminate against untimely protons. It would be relatively easy to measure the beam intensity in the single-turn abort line by standard techniques. The collimation system between the two dipoles in the transport line might also be instrumented. It would be desirable, albeit challenging, to continuously monitor the flux of untimely protons in the beam upstream of the production target. Veto tagging counters might be implemented as Cerenkov radiators viewed by photomultipliers. To provide the tremendous dynamic range needed, the photomultiplier gains might be modulated in two ways: by using Pockels cells as photon shutters, and by varying the voltage on one or more early stages of the dynode chain.

## 8.    EXPERIMENT LOCATION

The presently preferred location for the experimental area is Northwest of the Debuncher. The beam would be extracted at AP50, and an external beamline constructed downstream of the AP50 straight section (Fig. 8.1). The experiment would be located ~680 feet from AP50 in an area next to Giese Road (Fig. 8.2). This area is an undeveloped open meadow (i.e. not wetlands or woods) with easy access.

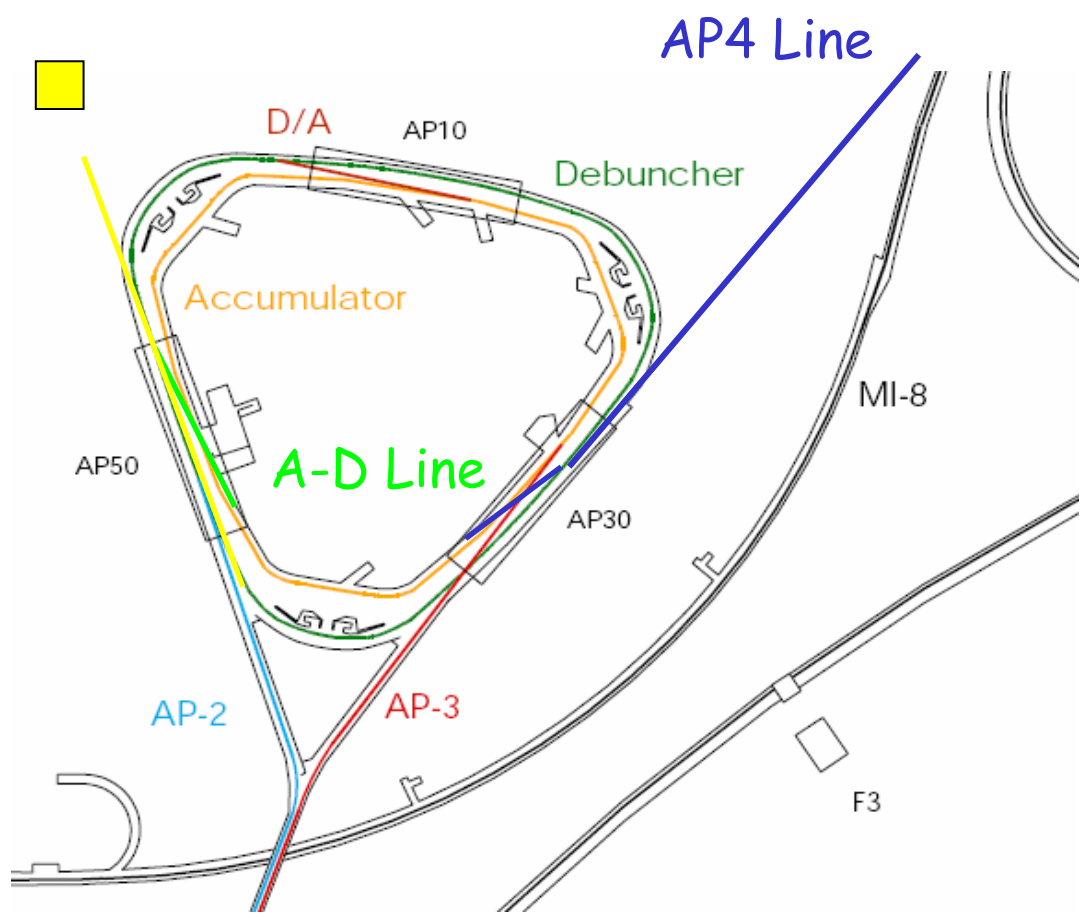

Figure 8.1: Preferred location for the beam extraction from the Debuncher is at AP50, with the experiment located to the northwest.

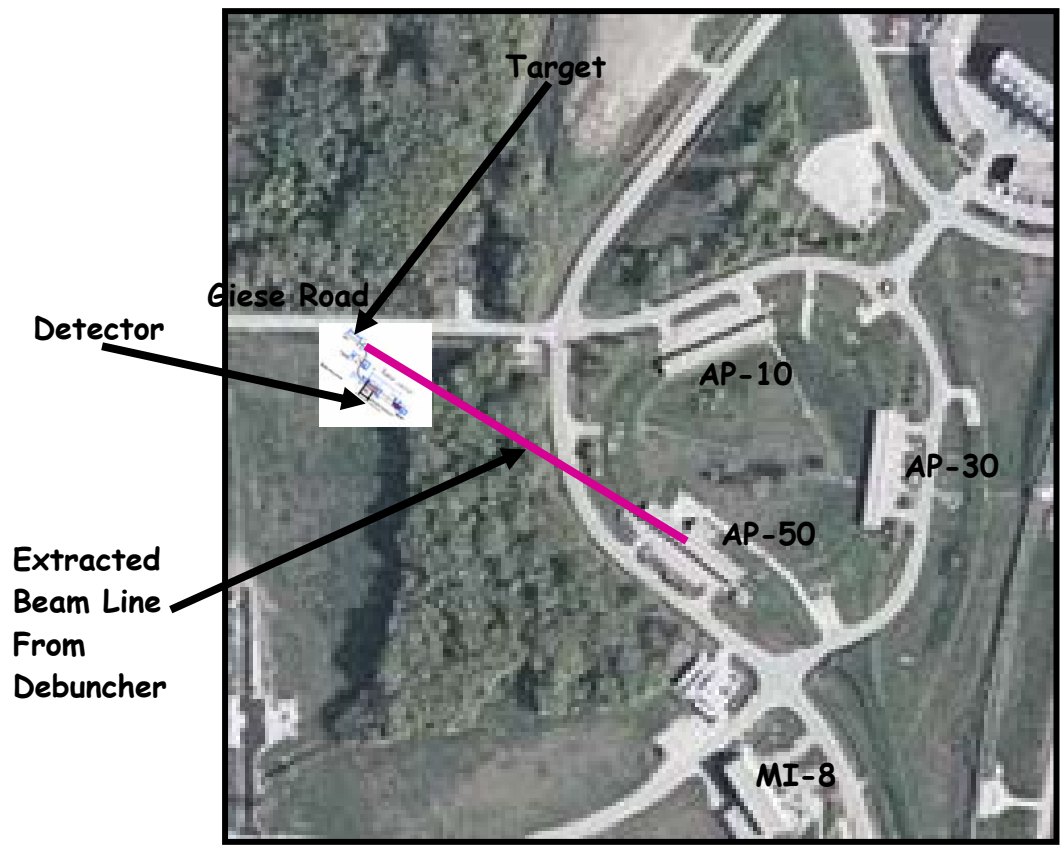

Figure 8.2: Preferred location for the experimental area. The white box contains the footprint for a MECO-like experiment, including proton target, decay channel, and detector.

## 9. SUMMARY

The SNuMI phase 2 upgrade to the proton source, if extended to include a modified Debuncher, would enable $10^{20}$ protons per year at 8 GeV to be rebunched and slowly extracted for a low energy muon program at Fermilab. This program could run in parallel with the future neutrino program, at the cost of reducing the number of MI protons for SNuMI by about 10%. To accomplish this will require, in addition to the foreseen SNuMI phase 2 upgrades, equipping the Debuncher with an rf system for the rebunching, reversing and relocating the D-A beamline so that protons can be transferred from the Accumulator to the Debuncher, implementing a slow extraction system in the Debuncher, and a beamline from the Debuncher to a new experimental area (for example, 680 feet to the northwest). The scheme described in this document has not been engineered, and there are many details that need to be better understood. Results from our initial studies are encouraging and motivate further work to develop the scheme. To ensure compatibility with the SNuMI program, it is desirable that this further work is integrated with the work on developing and planning the SNuMI phase 2 modifications.